\documentclass[conference]{IEEEtran}
\IEEEoverridecommandlockouts

\usepackage{cite}
\usepackage{amsmath,amssymb,amsfonts,mathtools}
\usepackage{algorithmic}
\usepackage[ruled]{algorithm2e}

\usepackage{graphicx,epsfig,color,graphics,caption}
\usepackage{subcaption}
\usepackage{epsf, psfrag, hhline, textcomp,enumitem}
\usepackage{xcolor}

\usepackage{times, fancyhdr}

\usepackage{multirow}
\usepackage{latexsym}
\usepackage{makecell}
\usepackage{comment}
\usepackage{tabularx} 

\usepackage{amsthm,amscd,amsbsy}

\usepackage[utf8]{inputenc}
\usepackage[english]{babel}
\usepackage[export]{adjustbox}
\usepackage{sidecap}
\usepackage{environ}

\def\BibTeX{{\rm B\kern-.05em{\sc i\kern-.025em b}\kern-.08em T\kern-.1667em\lower.7ex\hbox{E}\kern-.125emX}}

\newcommand{\mymarginpar}[1]{\marginpar{#1}}
\renewcommand{\marginpar}[1]{}

\newcommand{\ls}[1]{\dimen 0= \fontdimen 6 \the \font
	\lineskip =#1 \dimen0
	\advance \lineskip .5 \fontdimen 5 \the \font
	\advance \lineskip - \dimen 0
	\lineskiplimit =.9 \lineskip
	\baselineskip = \lineskip
	\advance \baselineskip \dimen 0
	\normallineskip \lineskip
	\normallineskiplimit \lineskiplimit
	\normalbaselineskip \baselineskip
	\ignorespaces
}

\newcommand{\bearn}{\begin{eqnarray*}}
	\newcommand{\eearn}{\end{eqnarray*}}
\newcommand{\barr}{\begin{array}}
	\newcommand{\earr}{\end{array}}

\newcommand{\N}{{\cal N}}

\newtheorem{definition}{Definition}
\newtheorem{assumption}[definition]{Assumption}
\newtheorem{property}[definition]{Property}
\newtheorem{proposition}[definition]{Proposition}
\newtheorem{lemma}[definition]{Lemma}
\newtheorem{theorem}[definition]{Theorem}
\newtheorem{corollary}[definition]{Corollary}
\newtheorem{example}[definition]{Example}
\newtheorem{remark}[definition]{Remark}
\newtheorem{axiom}[definition]{Axiom}

\newcommand{\benum}{\begin{enumerate}}
	\newcommand{\eenum}{\end{enumerate}}
\newcommand{\bdesc}{\begin{description}}
	\newcommand{\edesc}{\end{description}}

\newcommand{\bfig}[2]{\begin{figure}[htbp]\centering\includegraphics[width=#2]{#1}}
	\newcommand{\brotatefig}[2]{\begin{figure}[htbp]\centerline{\epsfig{figure={#1},clip=,angle=-90,width={#2}}}}	
		\newcommand{\bfigfirst}[2]{\begin{figure}[htbp]\centerline{\setlength{\epsfxsize}{#2}\epsffile{#1}}}
			\newcommand{\efig}[2]{\caption{#2}\label{fig:#1}\end{figure}\mymarginpar{fig:#1}}
		\newcommand{\erotatefig}[2]{\caption{#2}\label{fig:#1}\end{figure}\mymarginpar{fig:#1}}
	\newcommand{\rfig}[1]{Figure \ref{fig:#1}}
	\newcommand{\btab}[1]{\begin{table}\centering\begin{tabular}{#1}}
			\newcommand{\etab}[3]{\end{tabular}\caption[#3]{#2}\label{tab:#1}\end{table}\mymarginpar{tab:#1}\vspace{.1in}}
	
	\newcommand{\btabular}[1]{\begin{center}\begin{tabular}{#1}}
			\newcommand{\etabular}{\end{tabular}\end{center}}
	\newcommand{\bdefin}[1]{\begin{definition}\mymarginpar{def:#1}\label{def:#1}}
		\newcommand{\edefin}{\end{definition}}
	
	\newcommand{\bpro}[1]{\begin{property}\mymarginpar{pro:#1}\label{pro:#1}}
		\newcommand{\epro}{\end{property}}
	
	\newcommand{\bprop}[1]{\begin{proposition}\mymarginpar{prop:#1}\label{prop:#1}}
		\newcommand{\eprop}{\end{proposition}}
	
	\newcommand{\blem}[1]{\begin{lemma}\mymarginpar{lem:#1}\label{lem:#1}}
		\newcommand{\elem}{\end{lemma}}
	
	\newcommand{\bass}[1]{\begin{assumption}\mymarginpar{the:#1}\label{ass:#1}}
		\newcommand{\eass}{\end{assumption}}
	
	\newcommand{\bthe}[1]{\begin{theorem}\mymarginpar{the:#1}\label{the:#1}}
		\newcommand{\ethe}{\end{theorem}}
	\newcommand{\rthe}[1]{Theorem \ref{the:#1}}
	\newcommand{\bproof}{\noindent{\bfseries Proof.}}
	\newcommand{\eproof}{\hfill \squares \\ \vspace{.3cm}}
	\newcommand{\bcor}[1]{\begin{corollary}\mymarginpar{cor:#1}\label{cor:#1}}
		\newcommand{\ecor}{\end{corollary}}
	
	\newcommand{\bax}[1]{\begin{axiom}\mymarginpar{ax:#1}\label{ax:#1}}
		\newcommand{\eax}{\vspace{-.1in} \end{axiom}}
	
	\newcommand{\bex}[2]{\vspace{.1in}\begin{example}\mymarginpar{ex:#1}{\bfseries #2}\label{ex:#1} }
		\newcommand{\eex}{\end{example}\vspace{.3cm}}
	
	\newcommand{\brem}[1]{\begin{remark}\mymarginpar{rem:#1}\label{rem:#1}\em}
		\newcommand{\erem}{\end{remark}}
	
	\newcommand{\beq}[1]{\mymarginpar{eq:#1}\begin{equation}\label{eq:#1}}
	\newcommand{\beqno}[1]{\mymarginpar{eq:#1}\begin{eqnarray}\nonumber}
	\newcommand{\eeq}{\end{equation}}
	\newcommand{\eeqno}{&&\end{eqnarray}}
\newcommand{\req}[1]{(\ref{eq:#1})}

\newcommand{\bear}[1]{\mymarginpar{eq:#1}\begin{eqnarray}\label{eq:#1}}
\newcommand{\bearno}[1]{\mymarginpar{eq:#1}\begin{eqnarray}\nonumber}
\newcommand{\eear}{\end{eqnarray}}
\newcommand{\eearno}{\end{eqnarray}}
\NewEnviron{es}{
\begin{equation}\begin{split}
\BODY
\end{split}\end{equation}
}
\newcommand{\bieeeeq}[1]{\mymarginpar{eq:#1}\begin{IEEEeqnarray}{rCl}\label{eq:#1}}
\newcommand{\eieeeeq}{\end{IEEEeqnarray}}
\newcommand{\bsel}{\left\{\begin{array}{cl}}
\newcommand{\esel}{\end{array}\right.}
\newcommand{\bmat}[1]{\left[\begin{array}{#1}}
\newcommand{\emat}{\end{array}\right]}
\newcommand{\bsec}[2]{\mymarginpar{sec:#2}\section{#1}\label{sec:#2}}
\newcommand{\rsec}[1]{Section \ref{sec:#1}}

\newcommand{\bsubsec}[2]{\mymarginpar{subsec:#2}\subsection{#1}\label{subsec:#2}}
\newcommand{\rsubsec}[1]{Subsection \ref{subsec:#1}}

\newcommand{\bapp}{\begin{appendices}}
\newcommand{\eapp}{\end{appendices}}
\def\R{I\kern-0.30em R}
\def\N{I\kern-0.30em N}
\def\P{I\kern-0.30em P}
\newcommand \squares{\vrule height6pt width7pt depth1pt}

\def \ex{{\bf\sf E}}
\def \pr{{\bf\sf P}}

\newcommand \rezprob{r}

\newcommand{\pon}{p_{1,1}^{(i)}}
\newcommand{\poff}{p_{0,0}^{(i)}}

\begin{document}

\title{A Reinforcement Learning Approach for the Multichannel Rendezvous Problem}

\author{Jen-Hung~Wang, Ping-En~Lu, Cheng-Shang~Chang, and Duan-Shin~Lee\\
Institute of Communications Engineering\\
National Tsing Hua University \\
Hsinchu 30013, Taiwan, R.O.C. \\
Email: f0960778676@gmail.com; j94223@gmail.com; cschang@ee.nthu.edu.tw; lds@cs.nthu.edu.tw\\
}

\maketitle

\begin{abstract}
In this paper, we consider the multichannel rendezvous problem in cognitive radio networks (CRNs) where the probability that two users hopping on the same channel have a successful rendezvous is a function of channel states. The channel states are modelled by two-state Markov chains that have a good state and a bad state. These channel states are not observable by the users. For such a multichannel rendezvous problem, we are interested in finding the optimal policy to minimize the expected time-to-rendezvous (ETTR) among the class of {\em dynamic blind rendezvous policies}, i.e., at the $t^{th}$ time slot each user selects channel $i$ independently with probability $p_i(t)$, $i=1,2, \ldots, N$. By formulating such a multichannel rendezvous problem as an adversarial bandit problem, we propose using a reinforcement learning approach to learn the channel selection probabilities $p_i(t)$, $i=1,2, \ldots, N$. Our experimental results show that the reinforcement learning approach is very effective and yields comparable ETTRs when comparing to various approximation policies in the literature.
\end{abstract}


\bsec{Introduction}{labelofintroduction}
The multichannel rendezvous problem that asks two secondary users (SU) to find a common available channel (not used by primary users (PU)) is one of the fundamental problems in cognitive radio networks (CRNs) (see e.g., the book \cite{Book} and references therein). In view of possible jamming attacks \cite{Krunz2015}, the multichannel rendezvous problem is commonly solved by having each SU hopping on its available channels over time. When both SUs hop on a common available channel at the same time, it is assumed that a successful rendezvous occurs. For such a rendezvous problem, the objective is to minimize the time-to-rendezvous (TTR), i.e., the first time that the two SUs have a successful rendezvous. In the literature, there are various {\em deterministic} channel hopping (CH) sequences that can guarantee finite maximum time-to-rendezvous (MTTR) under various assumptions for CRNs, e.g., QCH \cite{Quorum}, DRSEQ \cite{DRSEQ}, Modular Clock \cite{Theis2011}, JS \cite{JS2011}, DRDS \cite{DRDS13}, FRCH \cite{ChangGY13}, ARCH \cite{ARCH}, CBH \cite{CBH2014}, and Two-prime Modular Clock \cite{ToN2017}. As pointed out in \cite{journalarxiv}, there is one practical factor that is not considered in these CH sequences, i.e., the channel states. Due to channel fading and interferences from other SUs, two SUs might not have a successful rendezvous even when they both hop on a common available channel at the same time. As such, it might be more practical to focus on the expected time-to-rendezvous (ETTR), instead of the MTTR.

In \cite{journalarxiv}, the authors considered a random channel state model, in which each channel has several random states and the probability that two SUs hopping on a common channel have a successful rendezvous is a function of the channel state. Specifically, for a CRN with $N$ channels, the states of the $N$ channels are characterized by a stochastic process $\{\boldsymbol{X}(t)=(X_1(t),X_2(t),\ldots,X_N(t)), t \ge 0\}$, where $X_i(t)$, $i=1,2, \ldots, N$, is the random variable that represents the state of channel $i$ at time $t$. The channel states are assumed to be not observable by a SU. When two SUs hopping on a channel in state $x$, they will rendezvous with probability $\rezprob(x)$. The authors in \cite{journalarxiv} considered the class of {\em blind rendezvous policies} in which each user selects channel $i$ {\em independently} with probability $p_i$, $i=1,2, \ldots, N$, in every time slot. They showed that there does not exist a channel selection policy (in terms of the channel selection probabilities, $p_i$, $i=1,2, \ldots, N$) that is universally optimal for any time-varying channel state model. For a fast time-varying channel model, the optimal policy is the single channel policy that only selects one particular channel. On the other hand, for a slow time-varying channel model, SUs should avoid selecting a single channel as that channel could be in a bad state for a long period of time.

Even though the channel states are not observable, one question is whether they can be implicitly learned (from either failed attempts or successful rendezvous) so as to speed up the rendezvous process in the future. To address such a question, we adopt a reinforcement learning approach to learn the channel selection probabilities of a SU. Reinforcement learning (see, e.g., the book \cite{learning} and the recent survey \cite{survey2019}) is a field of machine learning that addresses the problems of how to behave in an environment by performing certain actions and observing the reward from those actions. In these problems, the fixed limited resources must be allocated to maximize their expected gain. The reward of choice is only known at the time of allocation and may become better understood as time passes. Our problem is then to treat the channel selection probabilities as the fixed limited resources and learn how to allocate the channel selection probabilities to minimize ETTR. Specifically, our approach is to consider the multichannel rendezvous problem as the multi-armed bandit problem, and each successful rendezvous on a channel renders a reward for that channel. We then use the adversarial bandit algorithm, \textbf{Exp3}, in \cite{multiarmed} to learn the channel selection probabilities. When the $N$ channels are independent and identically distributed (i.i.d.), our numerical experiments show that \textbf{Exp3} yields comparable ETTRs to various approximation algorithms proposed in \cite{journalarxiv}. On the other hand, when channels are not i.i.d., \textbf{Exp3} is capable of learning the ``best'' channel. To the best of our knowledge, it seems that our paper is the first to study the multichannel rendezvous problem by a reinforcement learning approach.

\bsec{System model}{system}
\bsubsec{The multichannel rendezvous problem}{problem}
In this paper, we consider a cognitive radio network (CRN) with $N$ channels (with $N \geq 2$), indexed from $1$ to $N$, in the discrete-time setting where time is slotted and indexed from $t=0,1,2,\ldots$. We assume that there are two states for each channel, state 0 for the {\em bad} state and state 1 for the {\em good} state. Denote by $\rezprob(x)$ the rendezvous probability when a channel in state $x$, $x=0$ or 1. Then when two users hop on a channel in state $x$ at the same time, these two users will rendezvous with probability $\rezprob(x)$, and this is independent of everything else. Since state 0 is the bad state and state 1 is the good state, we assume that$$\rezprob(0) \le \rezprob(1) .$$

The states of the $N$ channels are characterized by the stochastic process $\{\boldsymbol{X}(t)=(X_1(t),X_2(t),\ldots,X_N(t)), t \ge 0\}$, where $X_i(t)$, $i=1,2, \ldots, N$, is the random variable that represents the state of channel $i$ at time $t$. The exact state of a channel at any time is not observable by a user. As discussed in \cite{journalarxiv}, the reason for that is because it is in general difficult for a user to know the congestion level of a channel (the number of users in a channel).

We consider the class of {\em dynamic blind rendezvous policies}, i.e., at the $t^{th}$ time slot each user selects channel $i$ with probability $p_i(t)$, $i=1,2, \ldots, N$. Such a channel selection is independent of everything else. Suppose that the channel state of the $i^{th}$ channel at time $t$ is $x_i$, $i=1,,2 \ldots, N$. Then under the dynamic blind rendezvous policy, the probability that these two users will have a successful rendezvous at time $t$ on channel $i$ is simply $(p_i(t))^2 \cdot \rezprob(x_i)$. This is because the two users have to hop on channel $i$ at time $t$ and the rendezvous is successful on channel $i$ with probability $\rezprob(x_i)$. As such, the two users will have a successful rendezvous at time $t$ is $\sum_{i=1}^N (p_i(t))^2 \cdot \rezprob(x_i)$. The objective is to learn a dynamic blind rendezvous policy (and the corresponding channel selection probabilities) that minimizes the expected time-to-rendezvous (ETTR).

\bsubsec{A Markov channel model with two states}{gtwo}
For the model of channel states, we consider the Markov chain with two states in \cite{journalarxiv}. We assume that the states of these $N$ channels are {\em independent}. The probability that the $i^{th}$ channel is in the good (resp. bad) state is $\rho_i$ (resp. $1-\rho_i$) for some $0 \le \rho_i \le 1$. As such, we have the following stationary joint distribution for the channel states
\bear{twojoint111g}
&&\pr (X_1(t)=x_1, X_2(t)=x_2, \ldots, X_N(t)=x_N) \nonumber\\
&&
=\prod_{i=1}^N \rho_i^{x_i} (1-\rho_i)^{1-x_i},
\eear
where $x_i$ (with the value being 0 or 1) is the state of channel $i$. For the $i^{th}$ channel, its channel state is characterized by a Markov chain with the transition probabilities:
\bear{tran1111}
&&\pr (X_i(t+1)=1 | X_i(t)=1)=\pon, \label{eq:tran1111aa}\\
&&\pr (X_i(t+1)=0 | X_i(t)=1)=1-\pon, \label{eq:tran1111ab}\\
&&\pr (X_i(t+1)=0 | X_i(t)=0)=\poff, \label{eq:tran1111bb} \\
&&\pr (X_i(t+1)=1 | X_i(t)=0)=1-\poff , \label{eq:tran1111ba}
\eear
where $0 < \pon, \poff <1$. Clearly, we have $$\rho_i=\pr( X_i(t)=1)=\frac{1- \poff }{(1-\pon) +(1-\poff)}.$$ Note that $$\mbox{Var}[X_i(t+1)]=\mbox{Var}[X_i(t)]=\rho_i (1- \rho_i)$$ and thus the correlation coefficient between $X_i(t+1)$ and $X_i(t)$, denoted by $\omega_i$, is
\bear{tran1155}
&&\frac{\ex [X_i(t+1) X_i(t)] -\ex[X_i(t+1)] \ex [X_i(t)]}{\sqrt{\mbox{Var}[X_i(t+1)] \mbox{Var}[X_i(t)] }}\nonumber\\
&&=\pon+\poff -1.
\eear
We say that the Markov chain $\{X_i(t), t \ge 0\}$ is positively correlated if $\omega_i \ge 0$. In this paper, we only consider positively correlated two-state Markov chains and we note the transition probabilities of the $i^{th}$ Markov chain can be characterized by the two parameters $\rho_i$ and $\omega_i$. It is shown in \cite{journalarxiv} that the ETTR of a blind rendezvous policy is bounded below when $\omega_i=0$ for all $i$ and it is bounded above when $\omega_i=1$ for all $i$. The argument used there can also be extended to show that the ETTR of a blind rendezvous policy is in fact increasing in $\omega_i$ when $\omega_i \ge 0$. Based on such structural results, various approximation algorithms for choosing the channel selection probabilities $p_1, p_2, \ldots, p_N$ of a (fixed) blind rendezvous policy were proposed in \cite{journalarxiv}. These policies include
\begin{description}
\item[(i)] Single selection policy: $p_1=1$ and $p_i=0$ for $i=2, \ldots, N$.
\item[(ii)] Uniform selection policy: $p_i=1/N$ for $i=1,2, \ldots, N$.
\item[(iii)] $(1+\epsilon)$-approximation policy: $p_i =\frac{\sqrt{u_i}}{\sum_{j=1}^N \sqrt{u_j}}$, where $u_1=1-(N-1)\delta$, $u_i=\delta$, $i=2, \ldots, N$, and $\delta=(\frac{\epsilon}{3(N-1)})^2$.
\item[(iv)] Harmonic selection policy: $p_i=c/i$, $i=1,2,..,N$, where $c$ is the normalization constant so that the sum of $p_i$'s is 1.
\item[(v)] Square selection policy: $p_i=c/i^2$, $i=1,2,..,N$, where $c$ is the normalization constant so that the sum of $p_i$'s is 1.
\item[(vi)] Sqrt selection policy: $p_i=c/i^{1/2}$, $i=1,2,..,N$, where $c$ is the normalization constant so that the sum of $p_i$'s is 1.
\end{description}
These 6 blind rendezvous policies will serve as the benchmarks for the comparison with our reinforcement learning approach. In particular, it was shown in \cite{journalarxiv} that the $(1+\epsilon)$-approximation policy achieves an asymptotic $(1+\epsilon)$-approximation ratio in the setting where either $\rezprob(0) \rightarrow \rezprob(1)$ or $\rezprob(0) \rightarrow 0$ and $\omega_i=1$ for all $i$. In our experiments, we set $\epsilon=0.2$.

\bsec{Reinforcement learning}{RL}
In this section, we adopt a reinforcement learning approach to learn the channel selection probabilities so as to minimize the ETTR. It is assumed that each user cannot observe the channel states of the (hidden) Markov chain. This is similar to the multi-armed bandit problem where a gambler does not know the success probability of a slot machine. For this, we formulate the multichannel rendezvous problem as an adversarial bandit problem \cite{exp3} in which there are $N$ possible actions, indexed from $1,2, \ldots, N$, in each time slot. The $i^{th}$ action corresponds to the selection of the $i^{th}$ channel. When two SUs rendezvous, one unit of reward is given to both users. Otherwise, there is no reward for the two SUs. For such an adversarial bandit problem, a famous algorithm to choose actions is the \textbf{Exp3} algorithm \cite{exp3} (which stands for "\textbf{Exp}onential-weight algorithm for \textbf{Exp}loration and \textbf{Exp}loitation"). In Algorithm \ref{alg:bandit}, we show the detailed steps of the \textbf{Exp3} algorithm for the multichannel rendezvous problem.
\begin{algorithm}
	\caption{\textbf{Exp3} for the multichannel rendezvous problem}\label{alg:bandit}
	\KwIn{A real parameter $\gamma \in (0,1]$ and a time horizon $T$}
	\KwOut{The channel selection probabilities $p_i(t)$, $i=1,2, \ldots, N$ and $t=1,2, \ldots, T$.}
	\textbf{Initialization}: Set $w_i(1)=1\ \text{for}\ i=1,...,N.$\\
	\textbf{For each} $t=1,2,...,T$
	
	\noindent 1: Set $p_i(t)=(1-\gamma)\frac{w_i(t)}{\sum_{j=1}^N w_j(t)}+\frac{\gamma}{N},\ i=1,...,N$.
	
	\noindent 2: Select a channel $i_t$ randomly accordingly to the probabilities $p_1(t),...,p_N(t)$.
	
	\noindent 3: Receive reward $z_{i_t}=1$ if there is a successful rendezvous, and $z_{i_t}=0$ otherwise.
	
	\noindent 4: For $j=1,...,N$, set\\
	$$\hat{z_j}(t)=\left\{
	\begin{aligned}
	&z_j(t)/p_j(t)\ &\text{if}\; j=i_t,\\
	&0\ &\text{otherwise}.
	\end{aligned}
	\right.	
	$$
	
	\noindent 5: Set $w_i(t+1)=w_i(t) \exp(\gamma\hat{z_i}(t)/N)$.
\end{algorithm}

To see the intuition of Algorithm \ref{alg:bandit}, we note that there are two terms in the channel selection probabilities $p_i(t)'s$ in Step 1. These two terms represent two fundamental concepts of reinforcement learning, exploration, and exploration. The first term in $p_i(t)$ is the "exploitation" term that makes the ``best'' decision given the current information. The second terms in $p_i(t)$ is the "exploration" term that allows us to gather more information that might lead to better decisions. These two concepts are rather intuitive for the channel selection problem. The exploitation term leads to a ``good'' channel. On the other hand, as the channel might change its state in the next time slot, the exploration term allows us to find another good channel. The parameter $w_i(t)$ is the weight of the channel $i$ at time $t$ and they are set to be 1 at time 1. When a successful rendezvous occurs, both SUs receive one unit of reward. We do not give a penalty to a channel selected by an SU that does not lead to a successful rendezvous. Therefore, two SUs have the same weights for all $t$ and thus the same channel selection probabilities $p_i(t)$, $i=1,2, \ldots, N$ for all $t$. The weight update rules in Steps 4 and 5 are the softmax update \cite{softmax} that increases the weight of a channel with a successful rendezvous.

The reward in Step 3 of the original \textbf{Exp3} algorithm in \cite{exp3} is assigned by an adversary. As there are two SUs in the multichannel rendezvous problem, SU 2 can be viewed as the adversary of SU 1. Intuitively, one might think such an adversarial viewpoint might be used for deriving an upper bound on the expected weak regret (defined as the difference between the maximum accumulated reward and the accumulated reward from the \textbf{Exp3} algorithm) like Theorem 3.1 of \cite{exp3}. However, as the channel selection probabilities of these two SUs are coupled through Algorithm \ref{alg:bandit}, the rewards of these two SUs are not independent of each other and the analysis in Theorem 3.1 of \cite{exp3} cannot be directly applied.

Another insight of Algorithm \ref{alg:bandit} is to view it as a stochastic game \cite{Shapley1953}. If $\rezprob(0)=\rezprob(1)=1$, then it is clear that the single selection policy that selects channel 1 all the time is optimal as both users rendezvous in every time slot. Through the process of exploration and exploitation, one expects that Algorithm \ref{alg:bandit} converges to the channel selection probabilities with $p_1=(1-\gamma)+\gamma/N$, $p_i=\gamma/N$, $i=2, \ldots, N$. This is exactly the $(1+\epsilon)$-approximation policy (for some $\epsilon$ that is a function of $\gamma$) that achieves the $(1+\epsilon)$-approximation ratio. Such an intuitive observation will be further verified in our experiments in \rsec{exp}.

\bsec{Experimental results}{exp}
In this section, we report our experimental results. For Algorithm \ref{alg:bandit}, we set $\gamma=0.02$. If $\gamma$ is set to be very large, then the probability distribution will be similar to the uniform selection policy. On the other hand, if $\gamma$ is very small, then the update of $w_i(t)$ is very small and that leads to a very slow convergence of the algorithm.

In our first experiment, we consider a system of 16 independent two-state Markov channels with the same parameters, i.e., $N=16$. The rendezvous probability at state 0 (resp. state 1) is $r(0)=0.001$ (resp. $r(1)=1$). There are 9 parameter settings for the two-state Markov channel model, the steady state probability $\rho=0.1$, 0.5 and 0.9, and the correlation coefficient $\omega=$0.1, 0.5, and 0.9.

\begin{figure*}[tb]
	\begin{center}
		\begin{tabular}{p{0.28\textwidth}p{0.28\textwidth}p{0.28\textwidth}}
			\includegraphics[width=0.28\textwidth]{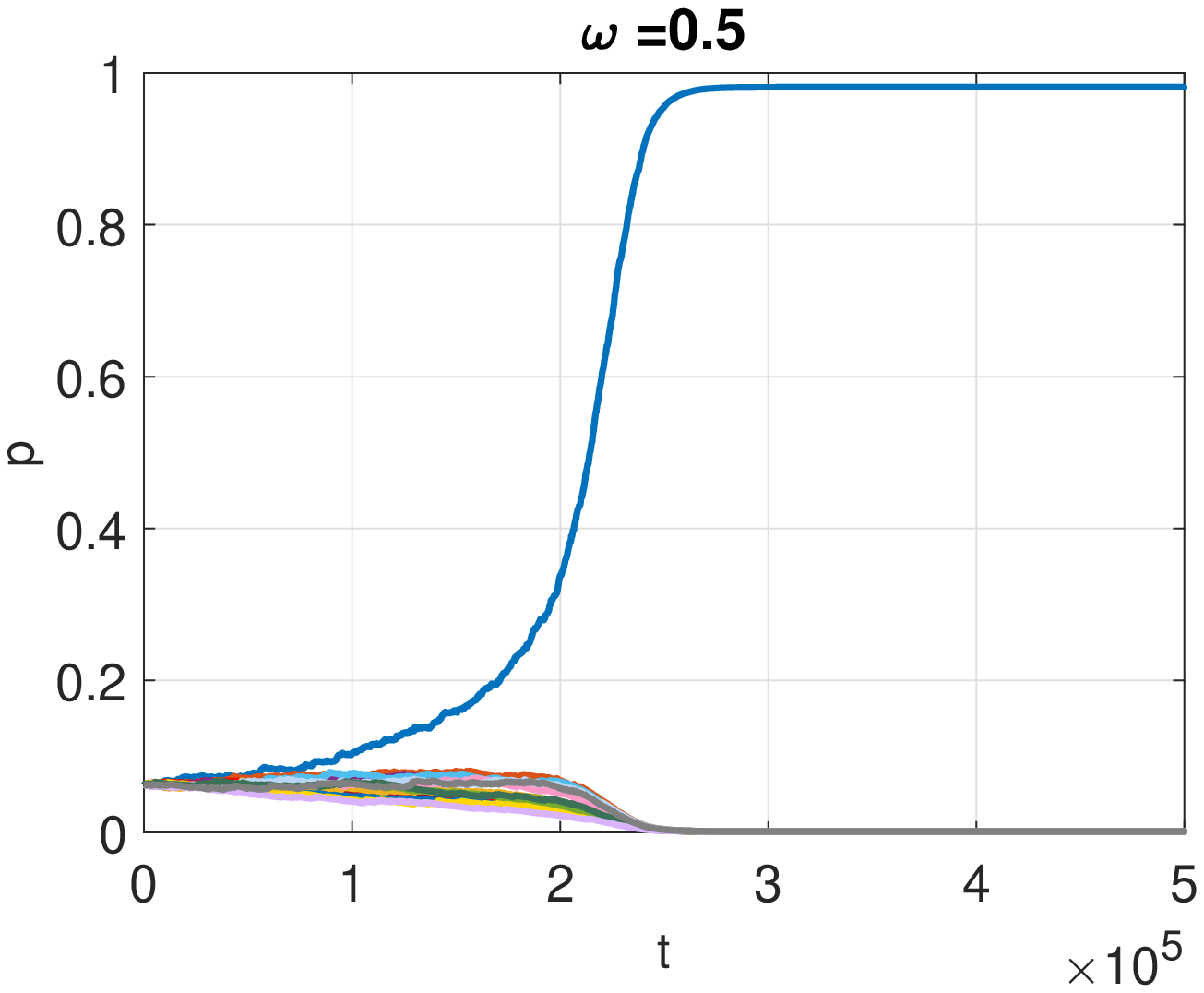} &
			\includegraphics[width=0.28\textwidth]{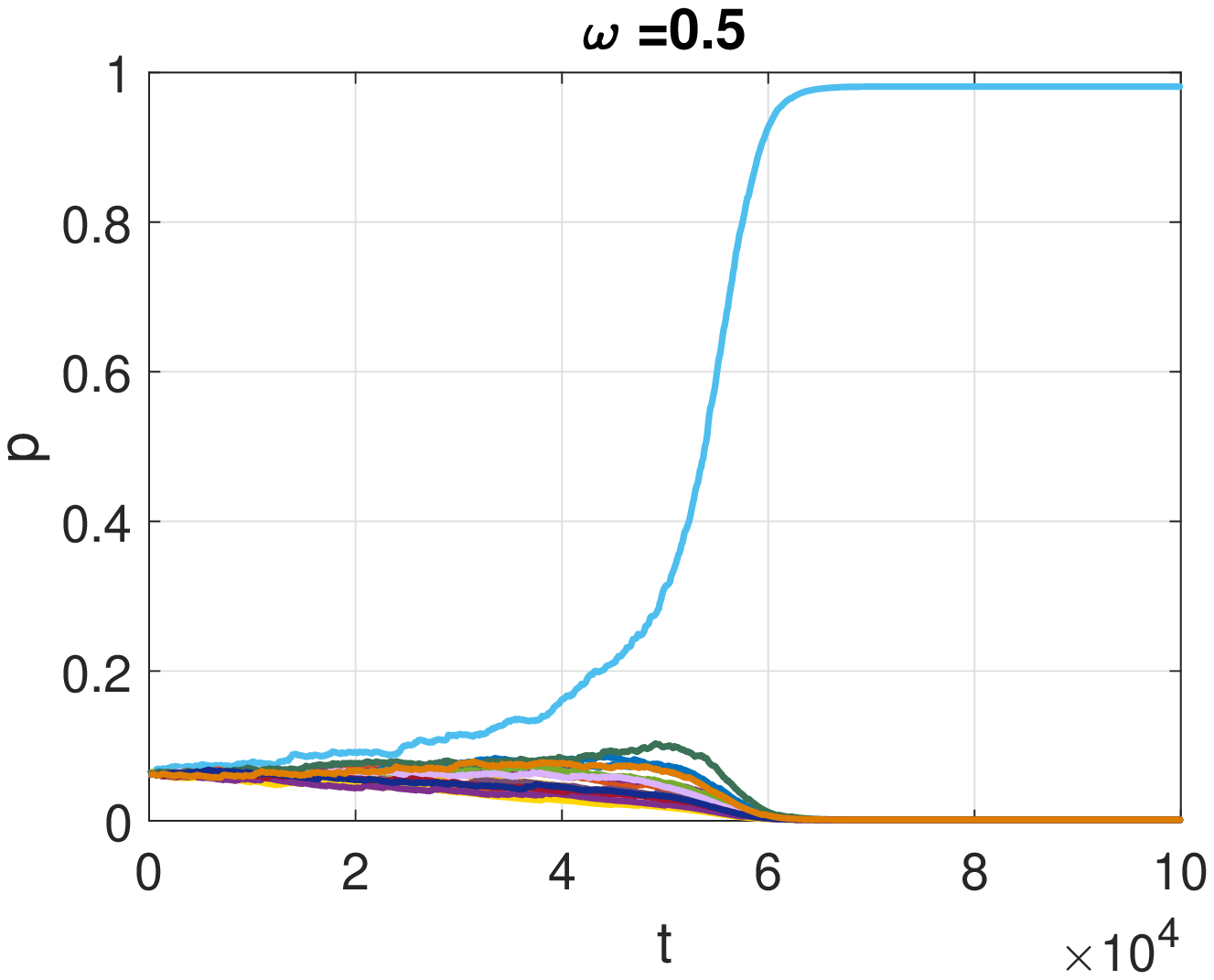} &
			\includegraphics[width=0.28\textwidth]{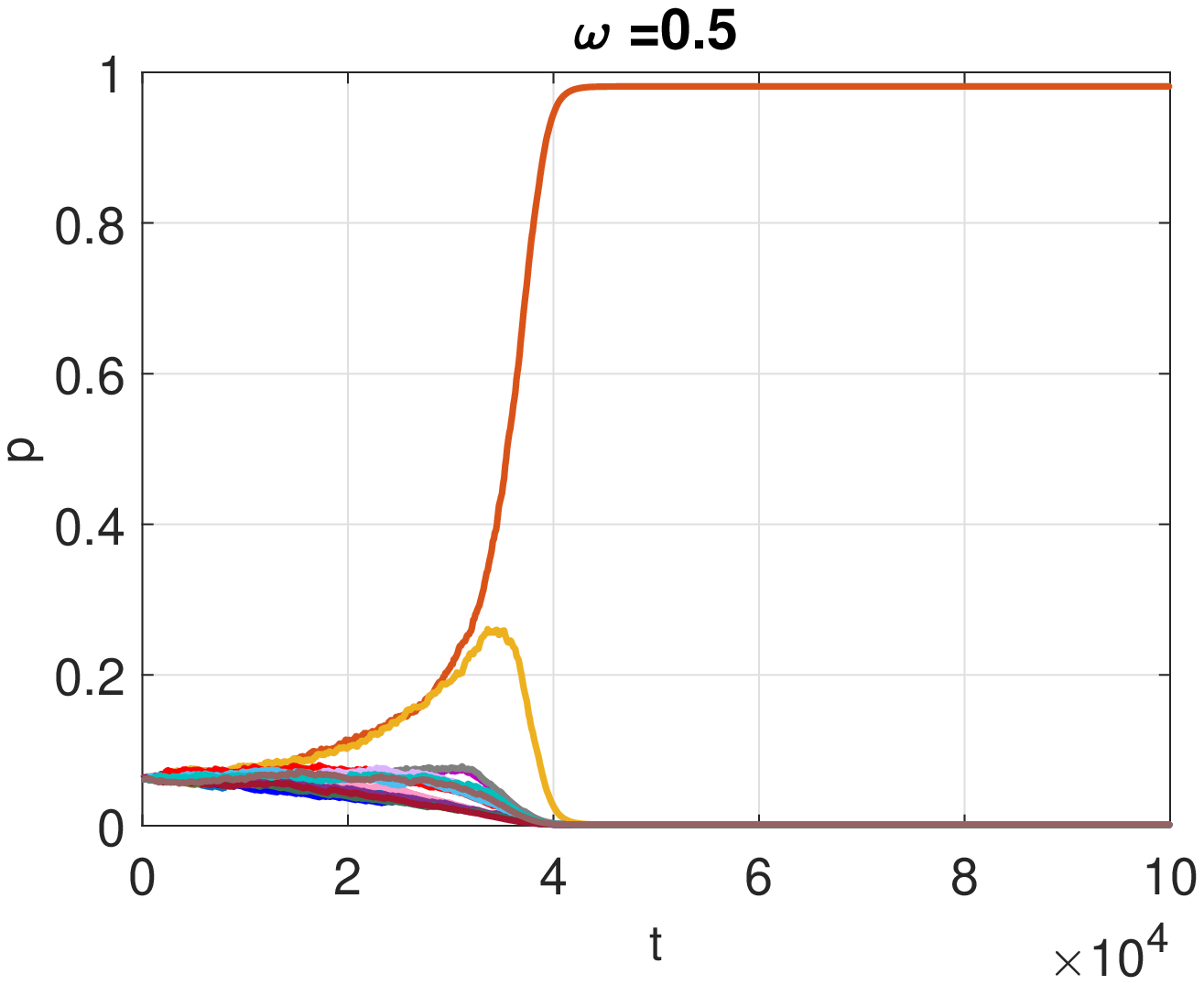}\\
			(a) $\omega=0.1$ & (b) $\omega=0.5$ & (c) $\omega=0.9$
		\end{tabular}
		\caption{The channel selection probability $p_i(t)$  with $\rho=0.1, 0.5, 0.9$ and $\omega=0.5$.}
		\label{fig:p_RL}
	\end{center}
\end{figure*}

For all the 9 settings, we find that the probability distributions learned by the algorithm when it converges are the same in every simulation. They all converge to the probability distribution $[0.98125, 0.00125, 0.00125, ...]$ (after sorting in the descending order of the channel selection probabilities). This is exactly the $(1+\epsilon)$-approximation policy in \cite{journalarxiv} with $\epsilon = 0.05732$. In \rfig{p_RL}, we plot the channel selection probability $p_i(t)$ with $\rho=0.1, 0.5, 0.9$ and $\omega=0.5$. Each curve (marked with various colors) in this figure corresponds to the channel selection probability of a channel with respect to time.

To see whether the \textbf{Exp3} algorithm converges to a good blind rendezvous policy, we measure the ETTR for the blind rendezvous policy with the channel selection probabilities with $p_1=0.98125$ and $p_i=0.00125$, $i=2, \ldots, 16$ and compare that with the 6 blind rendezvous policies described in \rsubsec{gtwo}. The ETTRs are obtained by averaging over 1000 independent runs for these blind rendezvous policies. In Table \ref{table:RL_ETTR_compare}, we show the comparison results for the 9 settings. For the fast time-varying channel model, i.e., $\omega \approx 0$, the optimal policy is the single channel policy \cite{journalarxiv}. For the slow time-varying channel model, i.e., $\omega \approx 1$, the $(1+\epsilon)$-approximation policy has the asymptotic approximation ratio $1+\epsilon$ \cite{journalarxiv}. From these numerical results, The ETTRs of \textbf{Exp3} are comparable to the best blind rendezvous policy among the 6 blind rendezvous policies described in \rsubsec{gtwo} for all the 9 settings.
\begin{table}[h]
	\centering
	\begin{subtable}{0.3\textwidth}
		\begin{tabular}{|l|l|l|l|}
			\hline
			\textbf{Policy$\backslash\omega$} & \textbf{0.1}     & \textbf{0.5}     & \textbf{0.9}     \\ \hline
			Single                 & 11.097  & 18.325  & 81.849  \\ \hline
			Uniform                & 156.968 & 156.007 & 159.818 \\ \hline
			Harmonic               & 74.290   & 79.734   & 100.212 \\ \hline
			$1+\epsilon$              & 12.041  & 19.865  & 92.220   \\ \hline
			Square                 & 23.572  & 29.714  & 81.369  \\ \hline
			Sqrt                   & 134.378 & 134.256 & 144.121 \\ \hline
			\textbf{Exp3}           		& 11.480   & 17.594   & 87.198  \\ \hline
		\end{tabular}
		\caption{$\rho=0.1$}
	\end{subtable}
	\begin{subtable}{0.3\textwidth}
		\begin{tabular}{|l|l|l|l|}
			\hline
			\textbf{Policy$\backslash\omega$} & \textbf{0.1}     & \textbf{0.5}     & \textbf{0.9}     \\ \hline
			Single                 & 2.089  & 2.884  & 10.724 \\ \hline
			Uniform                & 32.060  & 33.599 & 32.591 \\ \hline
			Harmonic               & 14.958 & 14.619 & 17.665 \\ \hline
			$1+\epsilon$              & 2.449  & 3.459  & 11.565 \\ \hline
			Square                 & 4.485  & 5.471  & 10.603 \\ \hline
			Sqrt                   & 25.062 & 26.952 & 27.184 \\ \hline
			\textbf{Exp3}            		& 2.282  & 2.957  & 10.616 \\ \hline
		\end{tabular}
		\caption{$\rho=0.5$}
	\end{subtable}
	\begin{subtable}{0.3\textwidth}
		\begin{tabular}{|l|l|l|l|}
			\hline
			\textbf{Policy$\backslash\omega$} & \textbf{0.1}     & \textbf{0.5}     & \textbf{0.9}     \\ \hline
			Single                 & 1.130   & 1.228  & 2.256  \\ \hline
			Uniform                & 17.994 & 17.477 & 17.515 \\ \hline
			Harmonic               & 7.894  & 7.727  & 8.271  \\ \hline
			$1+\epsilon$              & 1.280   & 1.368  & 2.150   \\ \hline
			Square                 & 2.735  & 2.661  & 3.280   \\ \hline
			Sqrt                   & 15.173 & 14.748 & 13.678 \\ \hline
			\textbf{Exp3}           		 & 1.148  & 1.265  & 2.249  \\ \hline
		\end{tabular}
		\caption{$\rho=0.9$}
	\end{subtable}
	\caption{Comparisons of the ETTRs of various blind rendezvous policies. }
	\label{table:RL_ETTR_compare}
\end{table}

\begin{figure*}[tb]
	\begin{center}
		\begin{tabular}{p{0.28\textwidth}p{0.28\textwidth}p{0.28\textwidth}}
			\includegraphics[width=0.28\textwidth]{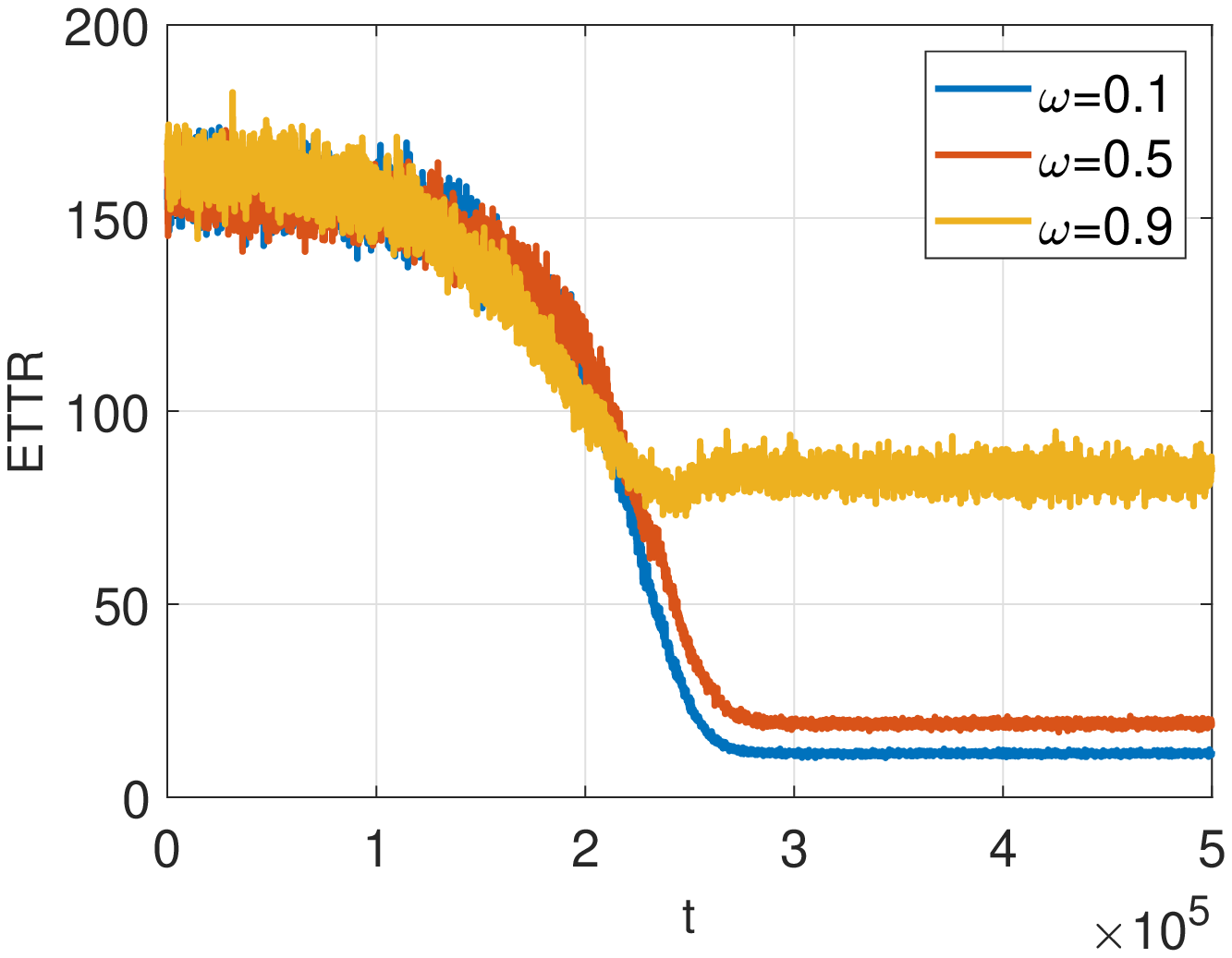} &
			\includegraphics[width=0.28\textwidth]{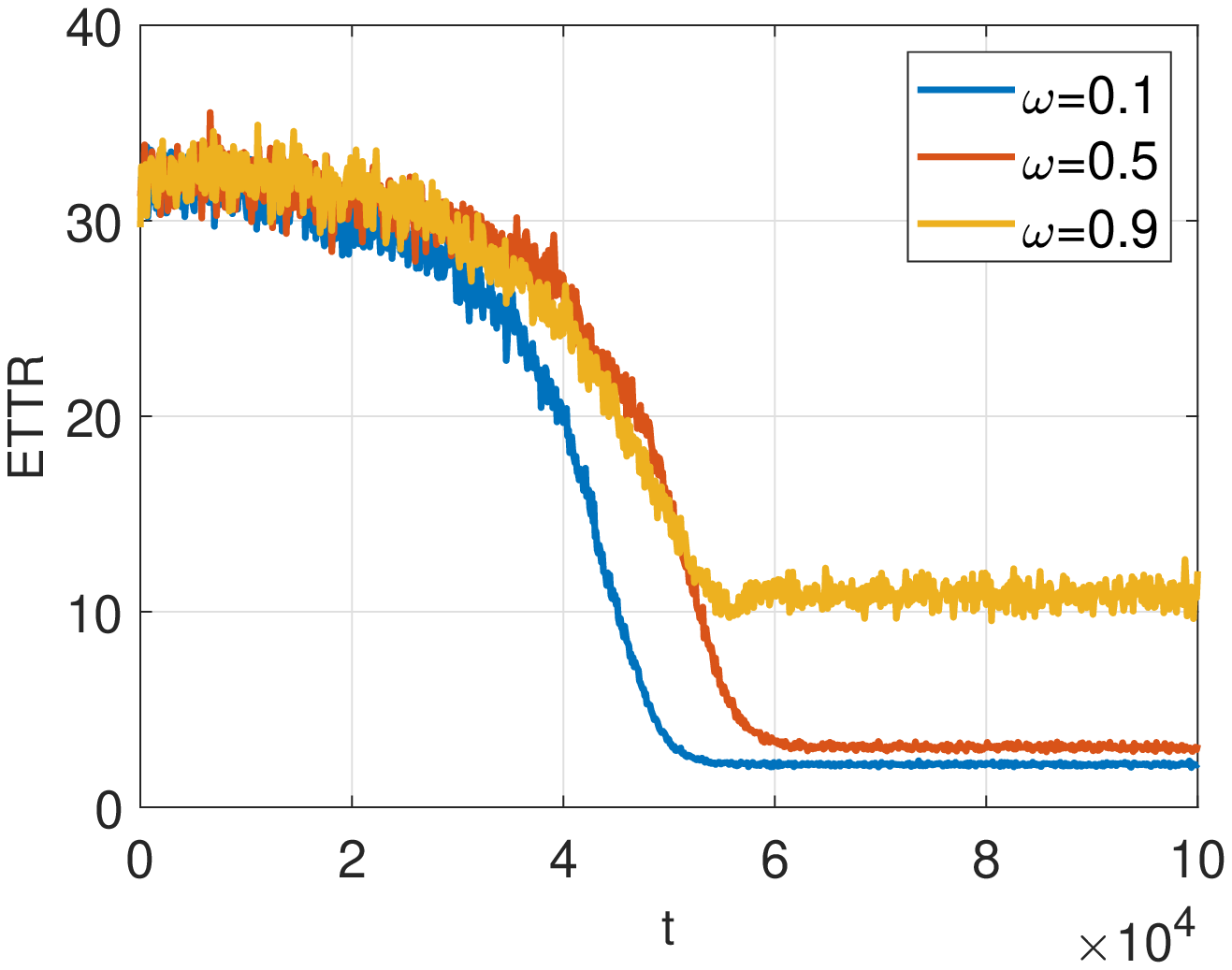} &
			\includegraphics[width=0.28\textwidth]{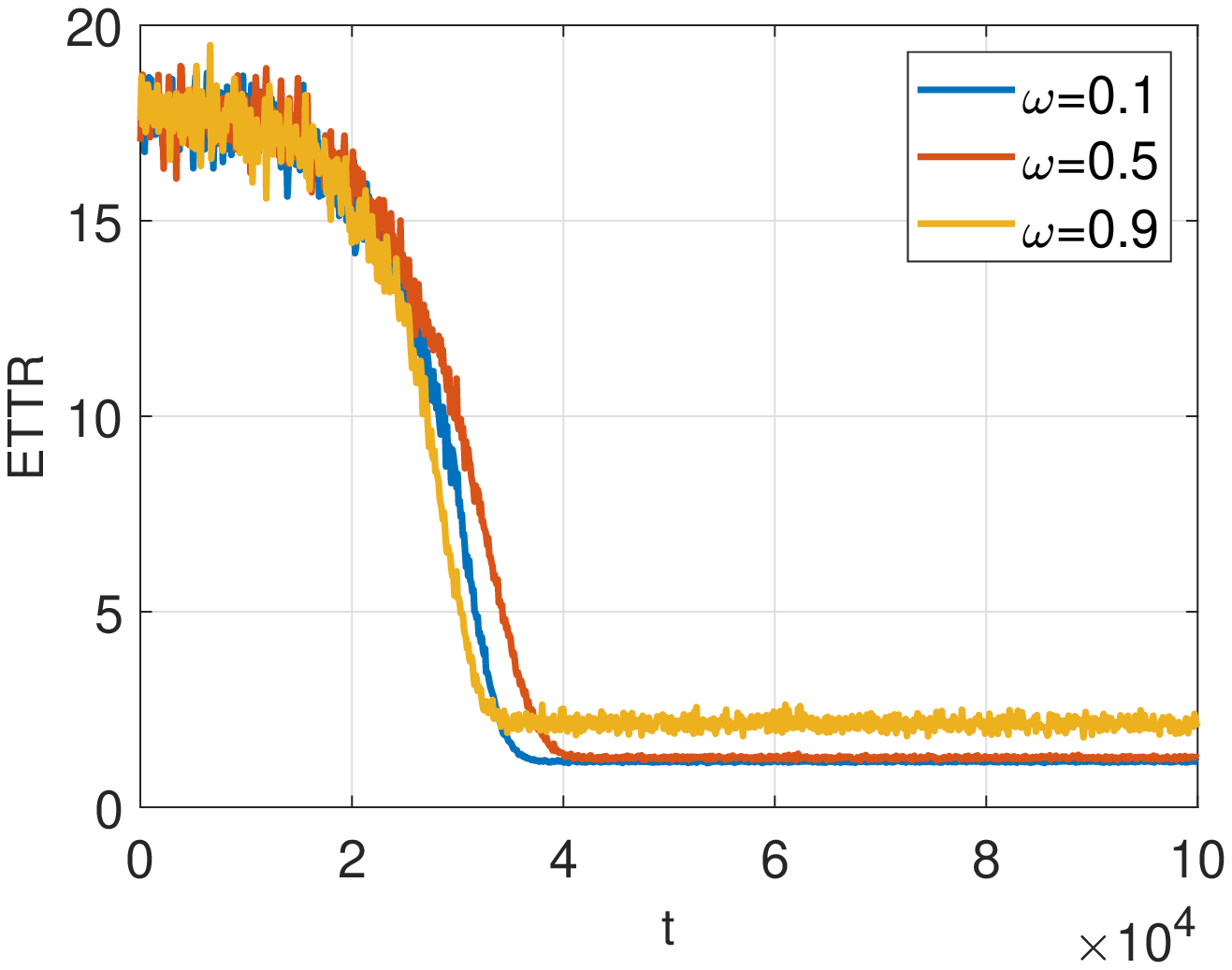}\\
			(a) $\omega=0.1$ & (b) $\omega=0.5$ & (c) $\omega=0.9$
		\end{tabular}
		\caption{The ETTRs (as a function of $t$) with $\rho=0.1, 0.5, 0.9$, respectively.}
		\label{fig:ETTR_RL}
	\end{center}
\end{figure*}

To show the effectiveness of Algorithm \ref{alg:bandit}, we measure the ETTRs for the channel selection probabilities $p_i(t)$, $i=1,2 \ldots, 16$ learned at time $t$ (by averaging over 1000 independent runs). In \rfig{ETTR_RL}, we plot the ETTRs as a function of $t$. As shown in this figure, all the ETTR curves are decreasing in time. This shows that Algorithm \ref{alg:bandit} is indeed learning better blind rendezvous policies with respect to time. One notable difference in these 9 settings is the convergence time of the algorithm. When $\rho$ is small, the probability that a channel is in a good state is also small. Hence, it is difficult to receive a reward for each channel. As such, it is more difficult to learn when $\rho$ is small and that leads to a longer convergence time. Moreover, we note that the fluctuation of ETTRs is much larger when $\omega=0.9$ (the yellow curves). The intuition behind this is that the channel state changes slowly when $\omega$ is large. But when a channel changes its state, it will take some time for the algorithm to learn such a change of states.

\begin{figure*}[tb]
	\begin{center}
		\begin{tabular}{p{0.28\textwidth}p{0.28\textwidth}p{0.28\textwidth}}
			\includegraphics[width=0.28\textwidth]{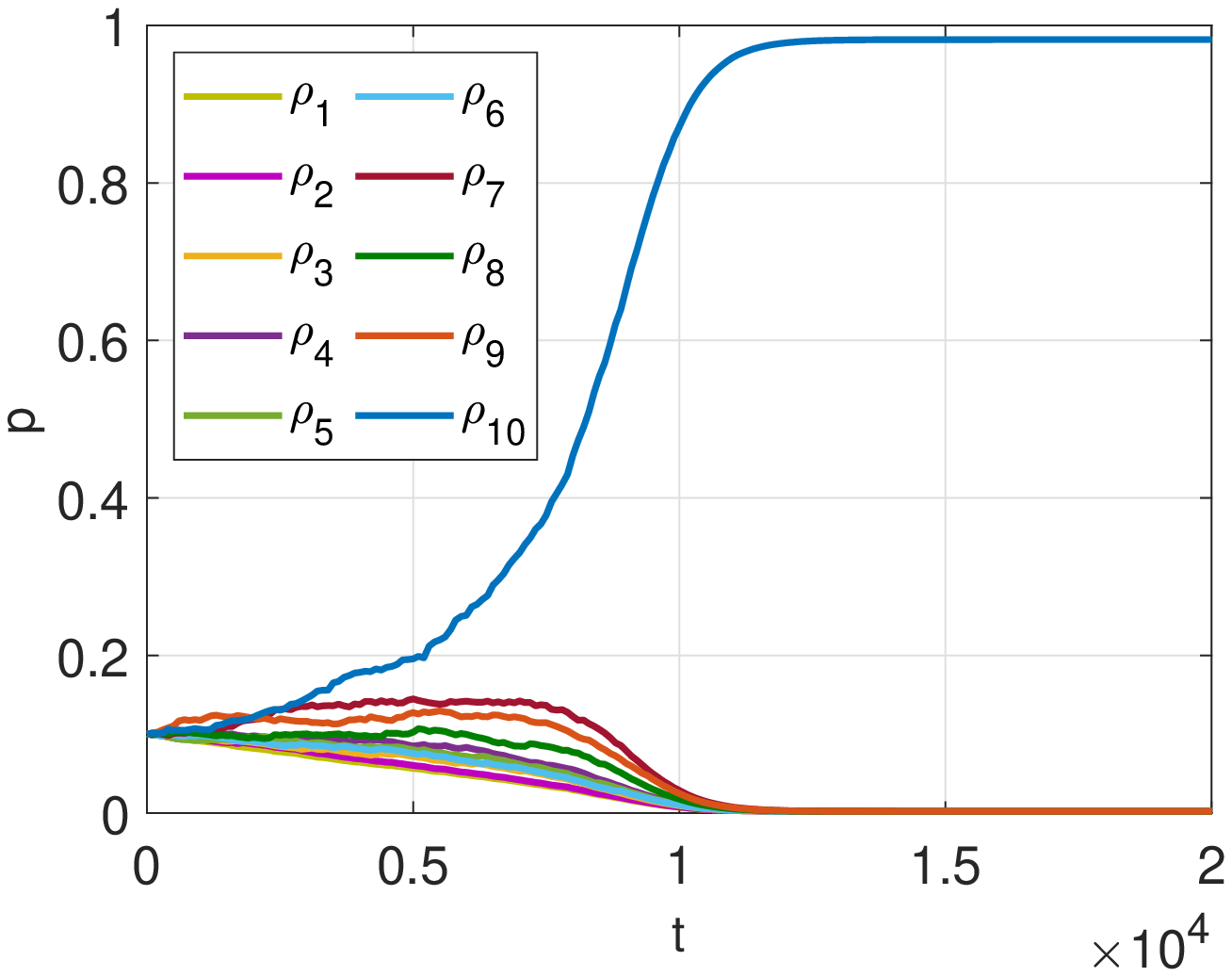} &
			\includegraphics[width=0.28\textwidth]{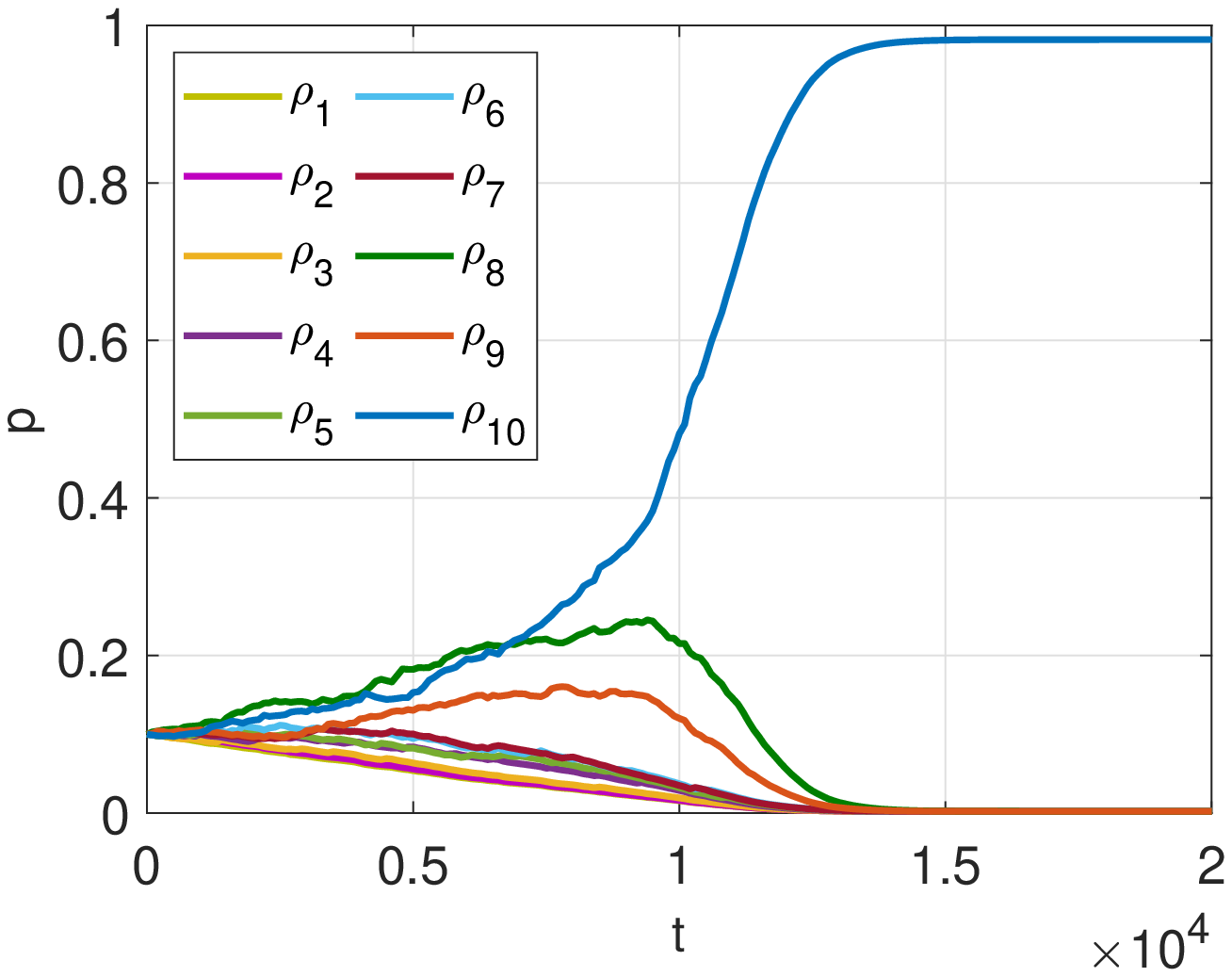} &
			\includegraphics[width=0.28\textwidth]{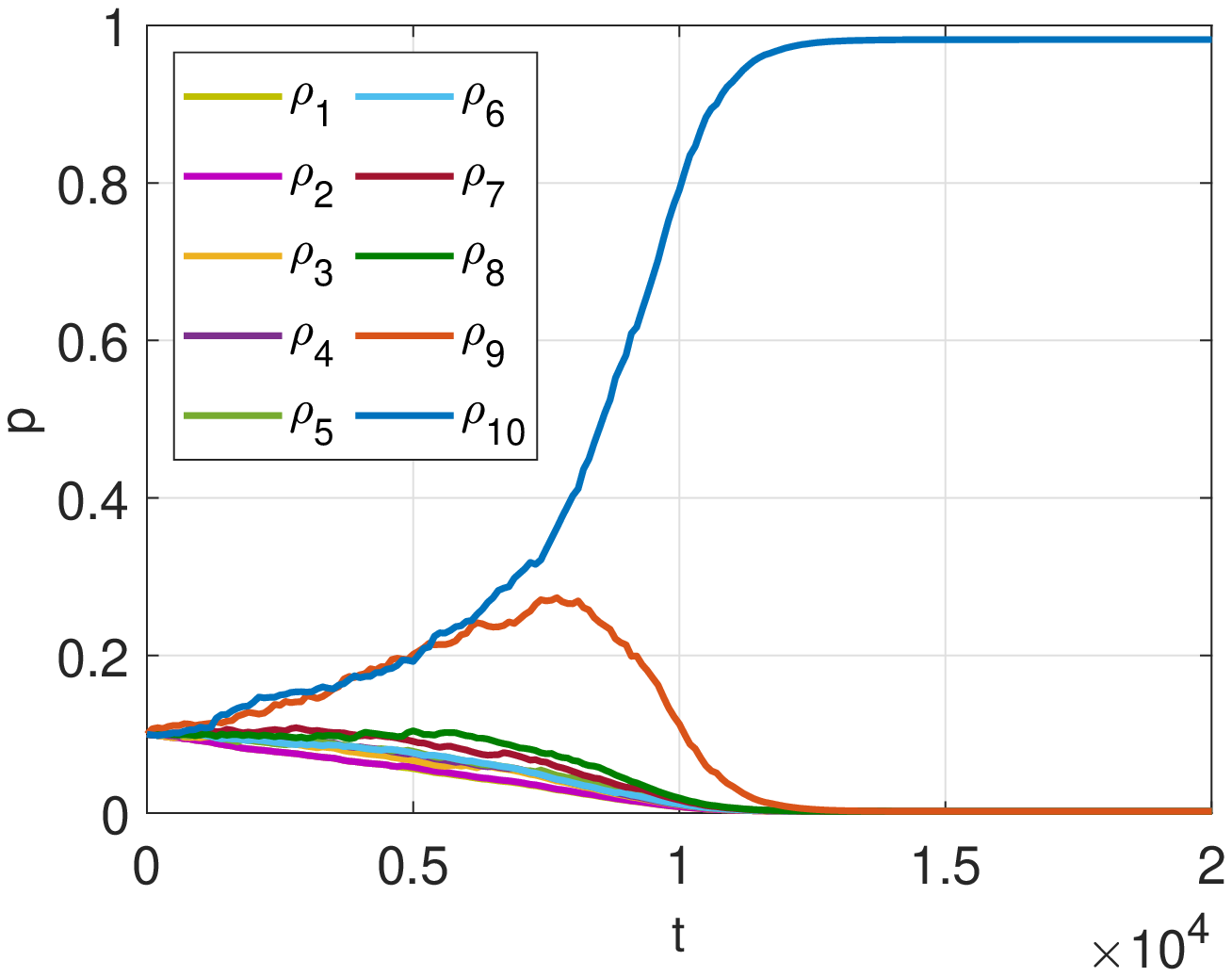}\\
			(a) $\omega=0.1$ & (b) $\omega=0.5$ & (c) $\omega=0.9$
		\end{tabular}
		\caption{The channel selection probability $p_i(t)$ with $\omega=0.1, 0.5, 0.9$, respectively.}
		\label{fig:diff_rho_RL}
	\end{center}
\end{figure*}

Even though the channel states are not directly observable, they can be implicitly learned. This is an additional advantage of Algorithm \ref{alg:bandit}. Instead of assuming that all the channels are identically distributed in \cite{journalarxiv}, we consider the setting with 10 channels that have different values of being in a good state. Specifically, We set $\rho_1=0,\rho_2=0.1,...,\rho_{10}=0.9$. In \rfig{diff_rho_RL}, we show the channel selection probability $p_i(t)$ with $\omega=0.1, 0.5, 0.9$, respectively. As shown in this figure, Algorithm \ref{alg:bandit} learns that channel 10 is the best channel in the long run and sticks to that channel with the probability 0.982. The channel selection probabilities for all the other channels are 0.002.

\bsec{Conclusion}{conclusion}
In this paper, we proposed a reinforcement learning approach for the multichannel rendezvous problem. When the channel states are not observable, we showed that the \textbf{Exp3} algorithm is very effective and yields comparable ETTRs when comparing to various approximation policies in the literature. One future work is to extend the reinforcement learning approach to the setting where the channel states are either observable or partially observable. In that setting, we need to develop effective $Q$-learning algorithms.


\end{document}